# Simultaneous Investigation of Ultrafast Structural Dynamics and Transient Electric Field by Sub-picosecond Electron Pulses

Run-Ze Li[1], Pengfei Zhu[1], Long Chen[1], Jie Chen[1,*], Jianming Cao[1,2], Zheng-Ming Sheng[1], Jie Zhang[1*]

[1]*Key Laboratory for Laser Plasmas (Ministry of Education) and Department of Physics and Astronomy, Shanghai Jiao Tong University, Shanghai 200240, China*

[2]*Physics Department and National High Magnetic Field Laboratory, Florida State University, Tallahassee, Florida 32310, USA*

[*]Correspondence: Jie Chen: jiec@sjtu.edu.cn; Jie Zhang: jzhang1@sjtu.edu.cn

**Abstract**

The ultrafast structure dynamics and surface transient electric field, which are concurrently induced by laser excited electrons of an aluminum nanofilm, have been investigated simultaneously by the same transmission electron diffraction patterns. These two processes are found to be significantly different and distinguishable by tracing the time dependent changes of electron diffraction and deflection angles, respectively. This study also provides a practical means to evaluate simultaneously the effect of transient electric field during the study of structural dynamics under low pump fluence by transmission ultrafast electron diffraction.

# 1. Introduction

Upon moderate femtosecond laser irradiation of a metallic sample, the optical energy is mainly absorbed and redistributed into the entire electron system by photon-electron interaction and electron-electron scattering. As illustrated in Figure 1, the laser-excited electrons are classified into two categories: "inward" and "outward" electrons. The inward energetic electrons remain inside the sample and initiate ultrafast structural dynamics of the lattice system through electron-phonon interaction [1] or transient perturbation to the interatomic potential [2]. The outward photoelectrons escape from the sample surface through thermionic and/or multiphoton emission and establish a surface transient electric field (TEF). The structural dynamics [3-6] and phase transition of solids [7-9] were directly measured with atomic spatiotemporal resolutions by ultrafast electron diffraction [10-14] or time-resolved x-ray diffraction/spectroscopy [15-18] and indirectly inferred by analyzing the electronic system behaviors obtained from time-resolved optical spectroscopy [19]. The surface TEF [20-22] and the associated magnetic field were mainly investigated by ultrashort electron and proton probes [23-27]. However, in most ultrafast electron diffraction (UED) experiments structural dynamics and transient surface electric fields are coexist and may be intertwined [22,28,29]. In some cases, the deflection of probe electrons produced by the transient surface electric field could be dominant and make it hard to interpret the recorded UED data[21,22,30]. At present, differentiation of these two effects in ultrafast diffraction experiments is an active research topic[14,31].

Using 0.6 ps electron pulses, we show here a simultaneous investigation of both ultrafast structural dynamics and surface transient electric field, concurrently generated by femtosecond laser interaction with a 25 nm aluminum film, from the same transmission electron diffraction patterns. The structural dynamics is extracted from the variation of the diffraction ring radii (diffraction angle changes), which shows the electron-phonon coupling and coherent phonon generation in the lattice system. The surface TEF is obtained from the shifting of the diffraction ring centroid (deflection angle changes), which shows a maximum field strength of ~$10^5$ kV/m above the sample surface building up in tens of picoseconds. Besides a comprehensive understanding of the transient processes induced by laser-excited electrons, this study also provides a practical approach to distinguish structural dynamics from the effect of TEF under low pump fluence in transmission ultrafast electron diffraction, which may help to improve its spatiotemporal resolution.

## 2. Experiments and Data Analysis

The ultrafast electron diffraction and deflection experimental setup is depicted in Figure 2. It consists of a 1 kHz Ti:sapphire femtosecond laser system, a 59 kV DC photoelectron gun, and a digital imaging system, which includes a phosphor screen, a micro channel plate (MCP) based image intensifier, and a lens-coupled charge-couple device (CCD) camera. The 1.0 mJ, 70 fs, 800 nm laser output was split into two parts. 90% of the output was used as the pump beam to initiate the transient processes. The remaining 10% was frequency tripled and directed to a photocathode with a 30 nm thick silver layer to generate the ultrashort electron pulse (probe beam) by photoelectric effect. After being extracted and accelerated up to 59 keV in 5 mm, the ultrafast electron pulse was shaped by a 40 μm pinhole, focused and collimated by a magnetic lens before probing the sample in a transmission configuration. The delay time (Δt) between the pump and probe pulses was set by a linear translation delay stage placed in the pump beam path. The sample, prepared by the standard procedure [3], was a freestanding 25 nm polycrystalline aluminum film with sub-micron domains confirmed by conventional Transmission Electron Microscopy, and it was mounted in an ultrahigh vacuum chamber with a base pressure better than $1.0 \times 10^{-9}$ Torr. In the experiments, the pump beam was focused onto the sample with a diameter of 1.2 mm ($1/e^2$ of the peak intensity), which was 12 times that of the probe electron beam to ensure a uniformly excited probing area. The pump was set at a low fluence of 2.1 mJ/cm$^2$ to avoid sample damage. On the average, each electron pulse was adjusted to contain ~ 2.5x10$^3$ electrons with a corresponding pulse duration of ~ 0.6 ps at the sample position [32]. Counting in the electron pulse width, the pump laser pulse width and the degradation due to the alignment of these two beams, a sub-picosecond overall temporal resolution was realized. The sample was tilted with a small angle ($\gamma=10^o$) and the electron diffraction pattern remained unchanged during the rotation. The electron diffraction pattern at each delay time, as such shown in Figure 2(c), was acquired with a ten-second integration by the digital imaging system. The four diffraction rings, from inner to outer side, correspond to the first-order diffraction profiles of (111), (200), (220) and (311) lattice planes of a face-centered cubic polycrystalline aluminum. The intensity centroid of (111) diffraction ring was taken as the center of all diffraction rings. After obtaining the coordinates of the diffraction ring centroid individually at each delay time, each 2D diffraction pattern was converted into a 1D diffraction intensity curve by radially accumulating the diffraction signal. Finally, each diffraction peak in the 1D curve was fitted with a Gaussian profile to derive the peak position (the diffraction ring radius). The surface TEF and structural changes were acquired simultaneously by analyzing the evolution of diffraction ring centroids (deflection angle changes) and radii (diffraction angle changes) as a function of delay time, respectively. To improve the signal-to-noise ratio (SNR), the data point presented at a given delay time was an average of more than 30 independent measurements.

## 3. Results and Discussions

Because of the resonance to a parallel band gap of aluminum [33], an efficient energy deposition of the 800 nm pump laser, about 14%, is achieved within the 7.4 nm optical penetration depth [34,35]. The optical energy is mainly deposited into the conduction electrons and then rapidly redistributed among the electron system by the strong electron-electron scattering and the ballistic motion of excited electrons across the 25 nm thick film. The electron system will reach a new thermal equilibrium state within a time scale comparable to the laser pulse duration of 70 fs. Afterwards, the evolution of excited electrons has two ways: moving inward and outward, Figure 1.

3.1. Structural dynamics induced by inward electrons:

The energy stored in the inward electron system is transferred to the lattice system via electron-phonon coupling and eventually the electrons and the lattice reach a new equilibrium temperature in a couple of picoseconds [1], a process widely described by the two temperature model [36,37]. The associated lattice motion was monitored by the change in the corresponding diffraction ring radius ($r$). According to the Bragg's diffraction formula, $2d\sin\theta = \lambda$, we have $\Delta\theta/\theta \approx \Delta\theta/\tan\theta = \Delta r/r = -\Delta d/d$ for small angle diffraction, where $d$ is the lattice plane spacing, $\theta$ is the diffraction angle and $\lambda$ is the wavelength of probe electrons. The relative change of the diffraction angle ($\Delta\theta/\theta$), as shown in Figure 3, was obtained by dividing the absolute radius change at each positive delay time over the average radius recorded before time zero, which is defined as the onset of the (111) diffraction angle change. The single exponential decay fitting of time dependent relative changes of Bragg diffraction angles in Figure 3 gives an time constant of $1.5 \pm 0.4$ ps, which is consistent with the electron-phonon equilibrium time in previous results [3].

The ultrafast electron heating combined with the sequent heating of lattice through electron-phonon coupling will launch a coherent lattice vibration that is normal to the film surface and centered at a new equilibrium position. This is the driving force for the observed oscillation of Bragg angles (oscillation of the lattice plane) displayed in Figure 3, which reaches its first minimum at about 5 ps after time zero, followed by several oscillations with a decreasing amplitude. This recorded lattice plane oscillations reveal a typical damped coherent phonon that can be generated in the ultrafast and homogeneous heating of a nano-film [38]. The ~9 ps oscillation period agrees well with the theoretical value estimated by the one-dimensional standing wave model [39] as $T = 2L/V$ ( $L$=25 ± 4 nm as the film thickness and $V$ =6420 m/s [40] as the sound velocity inside solid aluminum ).

3.2. Transient electric field induced by outward electrons

Owing to thermionic and multi-photon emissions, energetic electrons excited by femtosecond laser pulses could escape from the metallic surface and generate a surface TEF [20], Figure 1, which will deflect the negatively-charged probe electrons. In the transmission geometry employed here, the interplay between the electrons escaped from the sample and the positive residual charges on the sample eventually determined the overall behavior of the probe beam deflection. In general, the evolution of the charge distribution on the sample surface is a highly nonlinear and dynamical process involving the ejected electrons, film surface and bulk charges, and their time-dependent mutual interactions. Here, we use a simplified model to estimate the strength of the electric field in the following discussion. First, since the probe beam diameter is about 0.1 mm, it is assumed that the effect of sub-micron domain structures of the aluminum nanofilm on the TEF is negligible. Furthermore, considering that the pump laser diameter is on the order of millimeter, much larger than the probe electron beam diameter, a parallel-plate capacitor configuration was assumed to estimate the TEF. Therefore, the electric field formed between the positively-charged surface and the emitted electrons should be perpendicular to the sample surface after reaching an electrostatic equilibrium following the optical excitation.

The effect of the TEF is monitored by tracing the centroid evolution of probe electrons as shown in Figure 4. It was confirmed experimentally that the centroid shifting of the directly transmitted electrons (primary beam) is the same as that of the diffracted electrons within the limits of experimental error. Therefore, the centroid of entire probe electron beam is represented by the intensity centroid of diffraction rings. Here only the centroid evolution of the (111) ring is presented due to its stronger diffraction intensity than others'. The strength of TEF can be estimated by the deflection of probe beam centroid with respect to its original position before laser irradiation, $\Delta R$. Then, the absolute change of probe electron position is converted into the deflection angle through $\Delta \alpha = \Delta R / L$, where $L = 46\, cm$ is the distance between the sample and the detector (MCP screen). The components of TEF in X and Y directions can be estimated by tracing the probe beam deflection angles, $\Delta \alpha_x$ and $\Delta \alpha_y$, respectively. As shown in Figure 4, the trace of the probe beam centroid displays similar behaviors in both X and Y directions, therefore, the deflection in the X direction is chosen as a representative, which reaches its maximum around 126 ps. The averaged field component in the X direction, $E_x$, was calculated by $m\Delta V_x = qE_x t$ using a small angle approximation:

$$\Delta \alpha_x \approx \tan \Delta \alpha_x = \frac{\Delta V_x}{V_z} = \frac{qE_x t}{mV_z} = \frac{qE_x t}{m_e V_z / \sqrt{1 - V_z^2 / c^2}} \tag{1}$$

where $m_e$ and $q$ are the electron rest mass and charge, $m = m_e / \sqrt{1 - V_z^2/c^2}$ is the relativistic mass of electron, $c$ is the speed of light, $V_x$ and $V_z$ are the velocities of probe electrons along the X and Z directions, and $t$ is the interaction time between probe electrons and the TEF. Since the ejected electrons usually travel at a velocity on the order of 1 μm/ps [22] and the TEF reaches its maximum at about one hundred picoseconds, the distance experienced by the probe electrons is assumed to be around 100 μm for simplicity. Considering $V_z = 1.33 \times 10^8$ m/s for 59 keV probe electrons, and neglecting its changes by the TEF within 100 μm interaction range ( less than 10 eV in beam energy change assuming the field strength is on the order of $10^5$ V/m [24]), the interaction time $t$ is estimated to be around 0.75 ps. Therefore, the field strengths at the maximum deflections in the X and Y directions, 45 and 35 μrad, are estimated to be 49 and 38 kV/m by Eq. (1), respectively. Furthermore, for an approximate tilt angle of $\gamma = 10°$ between the surface normal and the probed beam travelling direction as shown in Figure 2(b), the maximum total electric field, $E_t$, is estimated to be $3.6 \times 10^5$ V/m by $E_t = \sqrt{E_x^2 + E_y^2} / \sin\gamma$, which is in consistence with previous theoretical predictions [20]. In addition, the temporal evolution of the TEF is visualized by the change of $\Delta\alpha_x$ as a function of delay time in the insert of Figure 4, which is found to build up and decay exponentially with the characteristic time constants of 37 ±2 ps and 137 ±6 ps, respectively.

3.3. Comparison of the influence of two processes

Photo-induced surface TEF commonly exists in ultrafast electron diffraction experiments. The onsets of the structural changes and the surface TEF were found at the same moment within the sub-picosecond experimental resolution, which implies that the two processes are generated concurrently. It was suggested that such TEF may have a large impact on the interpretation of structural dynamics for a silicon sample in an electron grazing-incident geometry [22]. The maximum field strength measured here for the aluminum thin film ($3.6 \times 10^5$ V/m) excited at 2.1 mJ/cm$^2$ is near one order magnitude higher than that for the silicon sample (38 kV/m) illuminated at 67.7 mJ/cm$^2$ [22]. As a result, the influence of the TEF is considerable and evidenced by the fact that it induces beam angle changes ($\Delta\alpha$) several times larger than those by structural dynamics ($2\Delta\theta$) as presented in Table 1. Therefore, the differentiation of these two types of angle changes becomes crucial in the measurement of structural dynamics using ultrashort electron pulses, especially for metals and alloys.

In the transmission geometry employed in this study, we demonstrated that the beam deflection induced by the TEF produces an overall shifting of the diffraction pattern, which can be corrected by tracing the centroid of the scattered electrons during the extraction of structural dynamics from the diffraction ring radii changes. Moreover, the structural dynamics was found to remain undisturbed by the surface TEF at the low pump fluence used in this study as supported by the following evidences: (1) the electron-phonon interaction time, 1.5 ps, is much shorter than the 37 ps formation and 137 ps decay time of the TEF, (2) the lattice plane oscillations are very distinct, and (3) the lattice plane spacing is found to stabilize at a new equilibrium position within 35 picoseconds shown in Figure 3, and remained unchanged up to 1.2 ns while the TEF continues to evolve. A comparison between the dynamics of the structural changes and the surface TEF is summarized in Table 1.

We have also studied the evolution of the freestanding polycrystalline aluminum sample with about twice the above mentioned pump fluence. The results indicated that the separation of the structural dynamics and transient electric field is still valid at pump fluence that is generally used [3,41,42] in the recoverable structural dynamics studies of freestanding nanometer thin samples.

This method might also be applied to study the interplay between lattice structure dynamics and surface TEF at much higher fluence, where the lattice plane spacing may be intervened by the surface TEF. However, at higher pump fluence, or when the size of the pump beam is comparable to that of the probe beam, the transient electric field effect might be more complicated and the separation of transient electric field and recoverable/unrecoverable structural dynamics may break down due to the additional distortion of diffraction patterns, which is inaccessible for this study due to the low damage threshold of the nanometer thin freestanding films.

In the previous ultrafast keV electron diffraction studies, the time resolved dynamics of some metals, semiconductors and 2D semimetals such as graphene, were mainly obtained from the changes of diffraction intensity [4,29,43,44] to eliminate the potential effects of the surface TEF on diffraction angle [30]. However, the evolutions of diffraction angle and intensity are characteristic for the crystallographic structural dynamics and energy transfer process, respectively, which are manifested also by their different temporal behaviors. The concurrent investigation presented here provides an applicative method to monitor the histograms of both diffraction angle and intensity in a transmission configuration, which may give an overall picture of the structural dynamics. Moreover, ultrafast relativistic electron diffraction has been developed recently with 3-5 MeV electron sources [45-47], which can penetrate deeper than keV electrons and promise to achieve sub-100 fs time resolution. With benefits offered by higher electron energy, new avenue would be opened for transmission diffraction studies of hundreds of nanometer thick samples that are inaccessible to its siblings in the keV range. Considering a bunch of 3 MeV electrons to probe the same structural dynamics and TEF presented in this study, the probe electron deflection angle induced by the TEF is on the order of 1 μrad, which is comparable to the diffraction angle change resulted from the structural dynamics, 1 μrad. Therefore, the effects of TEF should also be taken account of when analyzing MeV electron diffraction patterns in both transmission and reflection configurations.

## 4. Conclusion

In summary, the ultrafast structural dynamics and the surface transient electric field induced by a femtosecond laser pulse (70 fs, 800 nm) interacting with a 25 nm polycrystalline aluminum film were investigated simultaneously by 0.6 ps, 59 keV electron pulses in a transmission configuration. At a low excitation fluence, 2.1 mJ/cm$^2$, the electron-phonon coupling and coherent phonon generation revealed by structural changes form the structural evolution with atomic spatiotemporal resolutions, in the presence of the surface transient electric field that can reach hundreds of kV/m and lasts up to 1 ns. These two transient processes are found to be distinguishable from each other by their qualitatively different effects on the diffraction patterns. The simultaneous investigation also provides a method to evaluate the interplay between lattice structure dynamics and surface transient electric fields under low excitation fluences, which is critical in ultrafast electron diffraction and microscopy studies.

**Acknowledgements**

The authors would like to thank Mr. Xing Lin for aluminum deposition, and Mrs. Xin-Qiu Guo and Mr. Gang-Sheng Tong for Transmission Electron Microscopy inspection of the sample. This work was supported by the National Basic Research Program of China under Grant No. 2013CBA01500 and the National Natural Science Foundation of China under Grants No. 11004132, 11121504, 11327902 and 11304199.

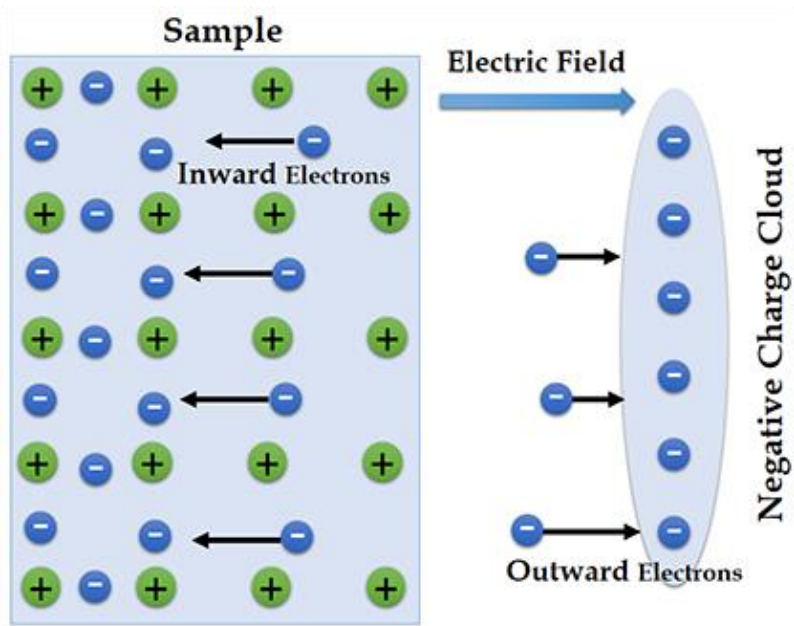

Figure 1. An illustration of the behaviors of inward and outward electrons generated by laser excitation of the electron system.

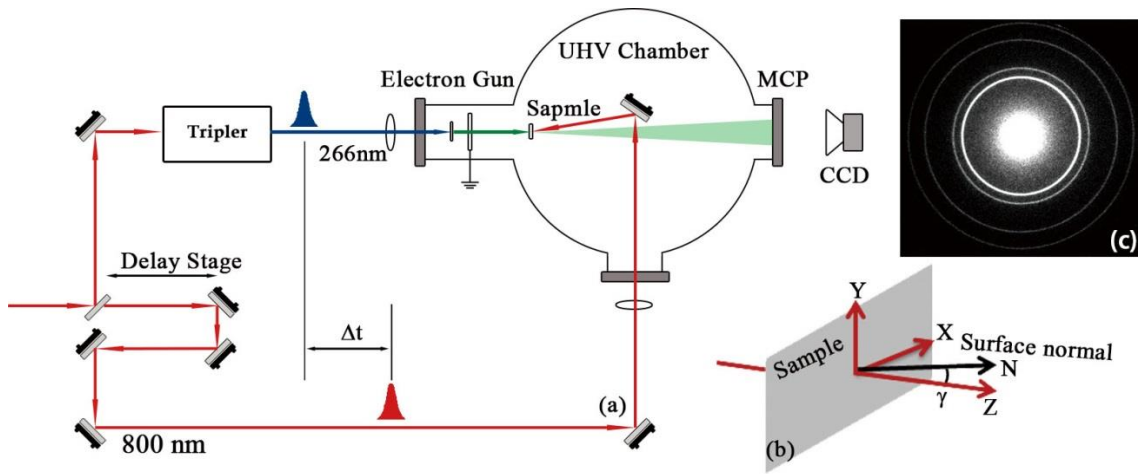

Figure 2. Schematic diagram of ultrafast electron diffraction and deflection. (a) Experimental configuration. (b) A more detailed view of the sample interaction area: Z denotes the probe electron traveling direction. The sample surface normal, which is parallel to probed transient electric field direction, has a small angle, $\gamma=10°$, with respect to the Z axis. X and Y denote the horizontal and vertical directions of CCD images and interpret the deflection directions of the probe electron beam. (c) A typical electron diffraction pattern of the 25 nm aluminum polycrystalline film collected with ~25 million electrons.

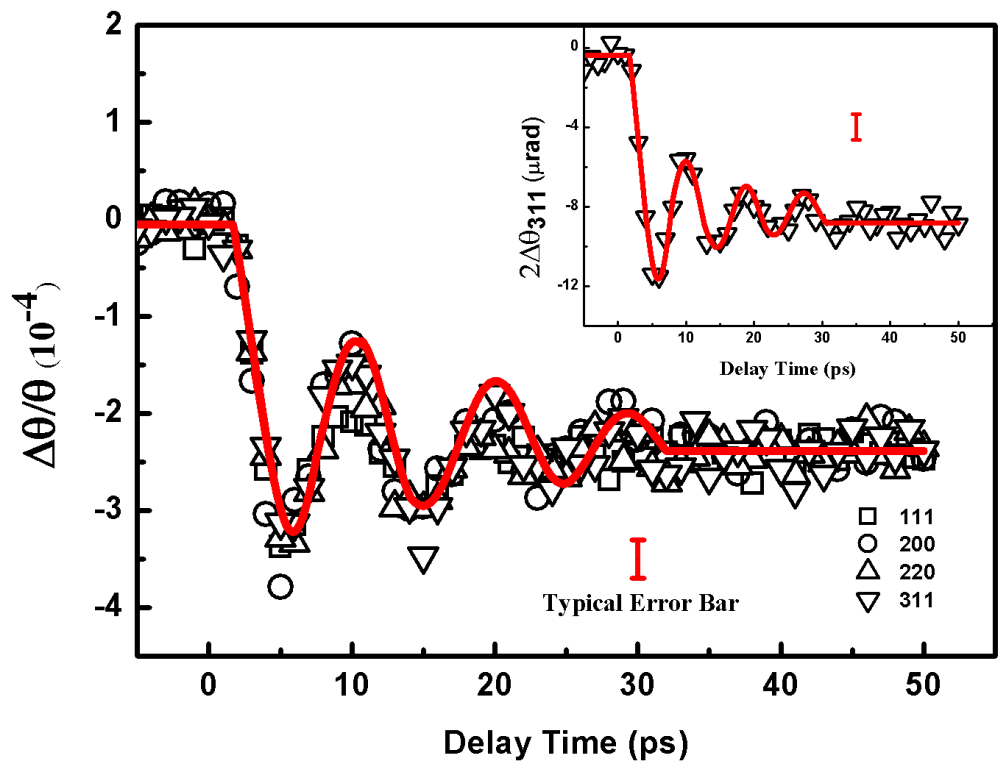

Figure 3. Time dependent relative changes of Bragg diffraction angles (Δθ/θ) for (111), (200), (220) and (311) lattice planes. Left insert: The detailed geometry of the sample surface normal N, the lattice plane normal $N_L$ and the probe electron beam path. Right insert: Time dependent changes of Bragg diffraction angle for (311) lattice plane ($2\Delta\theta_{311}$, twice of the change in diffraction angle). The solid curve is a fit to the data.

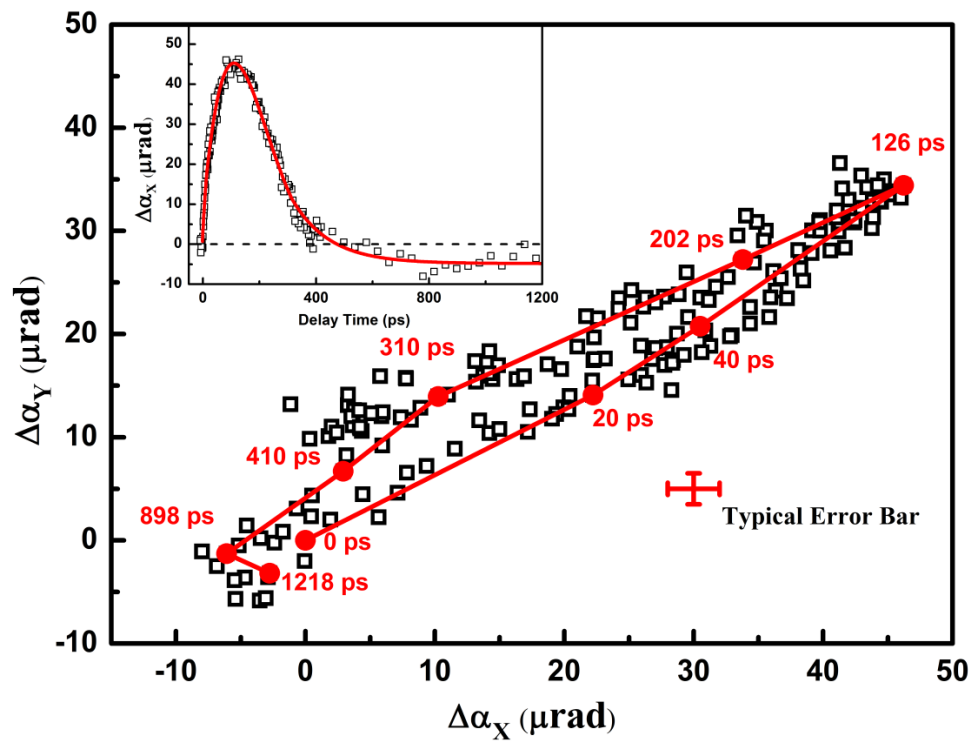

Figure 4. Time dependent trace of the (111) diffraction ring centroid. Several representative moments were marked by solid circles. The insert shows the time dependent deflection of the beam in the X direction, $\Delta\alpha_x$.

Table 1: A comparison between the dynamics of the structural changes ($2\Delta\theta_{311}$) and the surface transient electric field ($\Delta\alpha_x$). The damping time constant for coherent lattice oscillations is taken as the decay time for the structural changes.

|  | Build-Up Time (ps) | Decay Time (ps) | Maximum Change (μrad) |
|---|---|---|---|
| $\Delta\alpha_x$ | 37 | 137 | 45 |
| $2\Delta\theta_{311}$ | 1.5 | 14 | 12 |